\documentstyle[epsfig]{aipproc}
\def\etal{{\it et al.}}
\def\ginga{{\it Ginga}}

\begin{document}
\title{The Statistics of the BATSE Spectral Features}

\author{D.~Band$^{1}$, J.~Matteson$^{1}$, M.~Briggs$^{2}$,
W.~Paciesas$^{2}$, G.~Pendleton$^{2}$, and R.~Preece$^{2}$} 
\address{$^{1}$CASS, UC San Diego, La Jolla, CA 92093 \\
$^{2}$University of Alabama, Huntsville, AL 35899}

%\lefthead{LEFT head}
%\righthead{RIGHT head}
\maketitle

\begin{abstract}
The absence of a BATSE line detection during the mission's first six
years has led to a statistical analysis of the occurrence of lines in
the BATSE database; this statistical analysis will still be relevant
if lines are detected. We review our methodology, and present new
simulations of line detectability as a function of the line
parameters.  We also discuss the calculation of the number of
``trials'' in the BATSE database, which is necessary for our line
detection criteria. 
\end{abstract}

\section*{Introduction}
Whether spectral lines exist in the BATSE bursts is one of the most
pressing issues in burst spectroscopy since no BATSE detections have
been announced thus far\cite{palmer94,band96}.  The BATSE spectroscopy
team has been attacking this issue on many fronts:  searching for
lines\cite{briggs98}, analyzing the capabilities of BATSE's
spectroscopy detectors (SDs), checking that these detectors are
functioning correctly\cite{paciesas96}, and studying the statistics of
the detections and nondetections by BATSE and previous 
missions\cite{band94}.  Here
we describe our advances in these statistical studies. 
\section*{Line Statistics Formalism}
The line statistics methodology is built on a hierarchy of
probabilities based on the probability $p_i$ of detecting a line in a
spectrum\cite{band95}. In this discussion we use Roman subscripts 
to indicate spectra and Greek subscripts to denote bursts.  This detection
probability $p_i$ is primarily a function of the signal-to-noise 
ratio (SNR)
of the continuum and the angle between the detector normal and the
burst.  The probability is calculated through simulations.  
 
The probability $p_\alpha$ of detecting a line in a burst is a weighted 
sum of the
probabilities $p_i$ of detecting the line in each of the $N(N+1)/2$
possible consecutive spectra formed from the $N$ spectra accumulated
during the burst\cite{band97}. The weighting for each $p_i$ depends on
a model of how the line may occur within the burst (e.g., whether
lines are likely to persist a long or short time).  These
probabilities are specific to a given line type as parameterized
by the energy centroid,
line width and equivalent width. 
 
Clearly, this methodology requires information not only about the line
detections but also about all spectra in which lines were not
detected. Such extensive data are available for BATSE and {\it Ginga}.
Not coincidentally, {\it Ginga} reported the best documented line
detections. We have assembled a database of the necessary BATSE
information\cite{band97}, and we are collaborating with the {\it
Ginga} team in deriving the data for their detectors, as well as in
calculating the detection probabilities for the {\it Ginga} burst
detector.  Fenimore \etal\cite{fenimore93} calculated preliminary
detection probabilities for a {\it Ginga} line detection. 
 
The probabilities $p_\alpha$ of detecting a line in each of a given 
mission's bursts are then combined into the probability of the
observed pattern of detections and nondetections for that mission.
These
probabilities are calculated under certain assumptions (if only that
the detectors are modeled correctly), and the resulting probability  
is the likelihood for these assumptions. 

These likelihoods are used for various measures of the consistency
between BATSE and {\it Ginga}\cite{band94}.  For example, we developed
a Bayesian comparison of the consistency hypothesis (``lines exist and
both BATSE and {\it Ginga} are modeled correctly'') to various
alternative hypotheses that explain the apparent discrepancy (e.g.,
``BATSE is unable to detect lines'').  We have also developed a number
of standard ``frequentist'' (as opposed to Bayesian) consistency
measures such as the probability that if there are two detections,
both would be in the {\it Ginga} data.  Using a number of
approximations\cite{band97}, we find that currently BATSE and {\it
Ginga} are just marginally consistent; specifically, assuming that
lines exist and BATSE and {\it Ginga} are understood correctly, the
probability that {\it Ginga} would have the two line detections and BATSE
none is of order a few percent.  If lines exist, they are present in
only a few percent of all bursts.  These results are preliminary because
of the many approximations used (e.g., for the {\it Ginga}
detection probabilities). 
\section*{Line Simulations}
Our previous detectability studies only treated line sets with the
parameters of the reported \ginga\ detections\cite{band95}. However,
the \ginga\ lines may have been drawn from a distribution of lines
with different energy centroids, equivalent widths and intrinsic
widths.  In addition the computerized search is identifying emission
line candidates\cite{briggs98}. Therefore, we ran a large 
number of simulations of both absorption and emission lines with 
different line and observational parameters. 
The simulations used a photon model consisting of a continuum and a
spectral line.  The continuum was the canonical GRB functional
form\cite{band93} with $\alpha=-1$, $\beta=-2$ and $E_0=300$~keV.  We
used a multiplicative factor for the absorption lines and an additive
Gaussian line for the emission lines.  These lines were characterized
by the line centroid $E_{\rm cen}$ and the equivalent width $\Delta
E$.  The intrinsic line width is assumed to be less than the instrumental
resolution.
 
The resulting model photon spectrum was convolved with the response
matrix for a given burst angle $\theta$, and 200 realizations were
created by adding Gaussian noise appropriate to the count spectrum and
a representative background spectrum.  These realizations were then
fitted with continuum and continuum+line models, and the significance
of the line evaluated through the value of $\Delta \chi^2$ for 3 line
parameters (centroid, intrinsic width and equivalent width).  The strength 
of the continuum was measured by a normed SNR calculated over the
25--35~keV band; the SNR was varied by changing the accumulation time.
The detection probabilities are the fraction of these 200 simulations
whose significance exceeds a specific threshold, here $p(>\Delta
\chi^2)$ of $10^{-4}$ and $10^{-5}$. Simulations were done for
different values of the line centroid $E_{\rm cen}$, the equivalent
width $\Delta E$, the burst angle $\theta$, and the low energy cutoff
$E_{\rm low}$.  Figures 1 and 2 give examples of the dependencies on
$E_{\rm cen}$ and $\Delta E$. In these figures the y axes are 
the normed SNR at which there is a 50\% probability that the line 
would satisfy the significance threshold. 
\begin{figure}[b!] % fig 1
\centerline{\epsfig{file=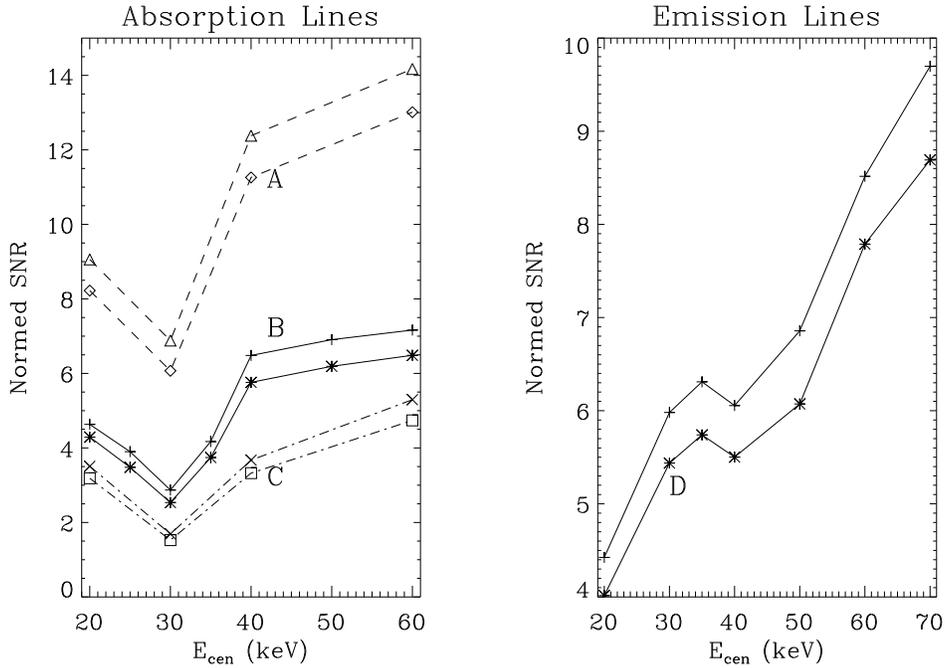,width=5.4in,height=3.6in}}
%\vspace{10pt}
\caption{Detectability as a function of the centroid energy $E_{\rm
cen}$.  The upper curves of each pair are for a significance threshold of
$10^{-5}$, and the lower for $10^{-4}$.  The y axes present the normed
SNR at which there is a 50\% probability that the line would satisfy
the significance threshold. 
The curves are identified by letters:  
A---$\Delta E=2.5$~keV, $\theta=30^\circ$, $E_{\rm low}=10$~keV;
B---$\Delta E=5$~keV, $\theta=30^\circ$, $E_{\rm low}=10$~keV;
C---$\Delta E=7.5$~keV, $\theta=30^\circ$, $E_{\rm low}=10$~keV;
and D---$\Delta E=5$~keV, $\theta=10^\circ$, $E_{\rm low}=10$~keV.
The dip in the absorption line curves and
the peak in the emission line curves around 30~keV results from the
line moving through the 25--35~keV band in which the SNR is measured.}
\label{fig1}
\end{figure}
\begin{figure}[b!] % fig 2
\centerline{\epsfig{file=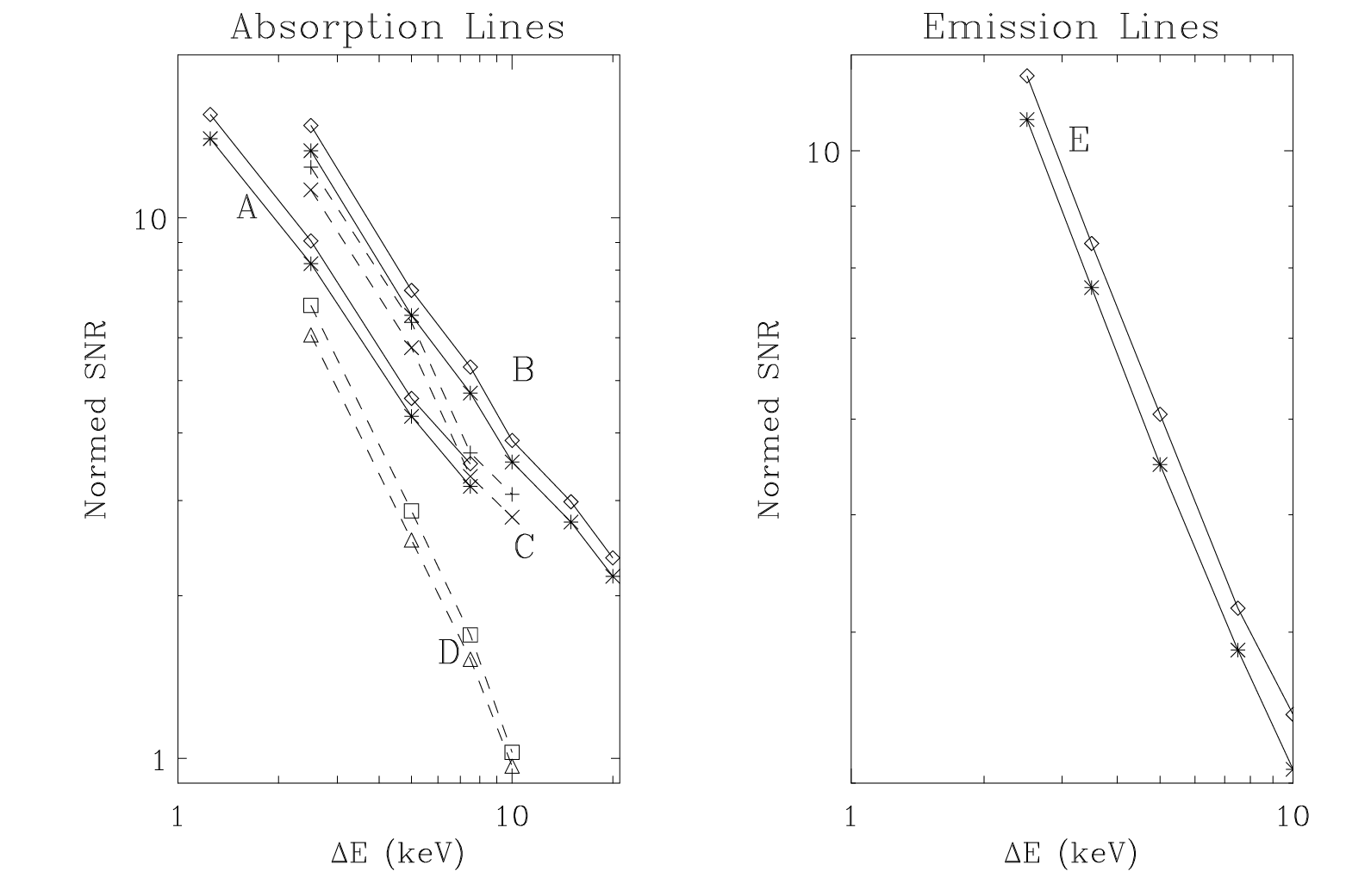,width=5.4in,height=3.6in}}
%\vspace{10pt}
\caption{Detectability as a function of equivalent width.  For every
parameter set there are 2 lines: the upper for a significance of
$10^{-5}$, and the lower for $10^{-4}$.  
The curves are identified by letters:  
A---$E_{\rm cen}=20$~keV, $\theta=20^\circ$, $E_{\rm low}=10$~keV;
B---$E_{\rm cen}=60$~keV, $\theta=30^\circ$, $E_{\rm low}=30$~keV;
C---$E_{\rm cen}=40$~keV, $\theta=30^\circ$, $E_{\rm low}=20$~keV;
D---$E_{\rm cen}=30$~keV, $\theta=30^\circ$, $E_{\rm low}=10$~keV;
and E---$E_{\rm cen}=40$~keV, $\theta=0^\circ$, $E_{\rm low}=10$~keV.
} 
\label{fig2}
\end{figure}
\section*{The Number of Trials}
The significance criterion must be set so that there is a small
probability of a false positive for the entire database.  The F-test
or the maximum likelihood ratio test provide the probability that a given
line feature is a fluctuation, but they do not include the number of
``trials,'' possibilities of obtaining that fluctuation. For example,
a line-like fluctuation could occur at a variety of energies with
different apparent line widths. Thus, while a given fluctuation may be
rare, if the number of trials is very large, then it might be probable
that such a fluctuation will occur somewhere within the burst database. 
 
The issue is how to calculate the number of trials.  The Bayesian
formalism for a feature's significance provides a conceptual
framework.  The Bayesian ``odds ratio,'' which compares the
probabilities for the continuum+line and continuum only models,
includes a factor that can be identified as the inverse of
the number of trials contributed by the line
parameters\cite{loredo92}. This ``Occam's
Razor'' factor suggests that the number of trials a parameter contributes 
is $\Delta A/\sqrt{2\pi} {\sigma_A}$, 
where $\Delta A$ is the range of possible values  and $\sigma_A$ is
the parameter's uncertainty.  This expression is strictly valid only
in the absence of any correlations among the parameters (e.g., when
the covariance matrices are diagonal). This dependence on the
parameter uncertainty is reasonable because parameter values separated
by less than the uncertainty are essentially \hbox{indistinguishable;} also
the uncertainty decreases and the number of trials increases as the
spectrum's SNR increases, as expected. 
 
Because the F-test or maximum likelihood ratio test are frequentist
tests and we derive the number of trials from a Bayesian expression,
it may be simpler to adopt the Bayesian methodology completely.   
Nonetheless, because the parameter uncertainty varies across the
spectrum, between spectra and from burst to burst, accounting for
the number of trials results in 
a complicated multiple integral over the various parameters. 
\section*{Acknowledgments} 
The work of the UCSD group is supported by the {\it CGRO} guest
investigator program and NASA contract NAS8-36081.

\end{document}